# ANALYSIS AND INTERFACE FOR INSTRUCTIONAL VIDEO

*Alexander Haubold and John R. Kender*

Department of Computer Science, Columbia University, New York, NY 10027

{ah297,jrk}@columbia.edu

## ABSTRACT

We present a new method for segmenting, and a new user interface for indexing and visualizing, the semantic content of extended instructional videos. Using various visual filters, key frames are first assigned a media type (board, class, computer, illustration, podium, and sheet). Key frames of media type board and sheet are then clustered based on contents via an algorithm with near-linear cost. A novel user interface, the result of two user studies, displays related topics using icons linked topologically, allowing users to quickly locate semantically related portions of the video. We analyze the accuracy of the segmentation tool on 17 instructional videos, each of which is from 75 to 150 minutes in duration (a total of 40 hours); it exceeds 96%.

## 1. INTRODUCTION

Video segmentation, indexing, and visualization are essential parts of content-based video retrieval. Characteristically of their genre, instructional videos tend to be taken in a set environment with a small set of well-defined areas of interest. The segmentation process should exploit this underlying structure.

Most related work has focused on indexing methods for news videos [1, 2], sports videos [3], and situation comedies [4]. Some work has been done on the segmentation of the blackboard frames of instructional videos taken in a specially instrumented classroom [5]. However, many instructional videos contain material from sources other than the blackboard, and from environments not specifically designed for video analysis.

The summarization and indexing of a video begins by collecting key frame images taken at points of substantial change. In our application, consisting of 17 videos of long lectures, these key frames were chosen by a proprietary software product sensitive to image motion. On average, a key frame is produced every 20 to 25 seconds, so 75 (150) minute lectures contain about 200 (350) key frames. However, they tend to be rather repetitive, as much lecturing consists in emphasizing verbally what has been

Acknowledgements: This work was supported in part by NSF grant EIA-00-71954 and an equipment grant from Microsoft.

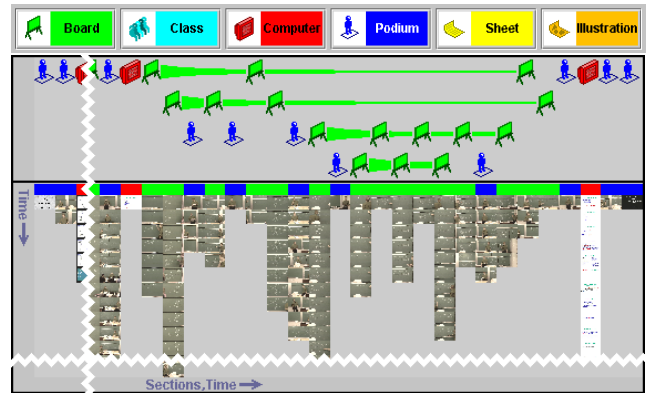

**Figure 1.** Top level of user interface: topological index with key frame summary. (A) above: Each key frame media type is assigned a distinguishable color as well as a descriptive icon. (B) below: Vertical key frame summaries are aligned with media type icons; horizontal topological groupings capture topic commonalities. Icons and key frames are clickable; they select topics, magnify the thumbnails, and pop up the video at the appropriate frame

visually created. By grouping together key frames of similar contents into topic clusters, the complexity of the key frame set can be reduced by 80 to 95%.

Structured experiments involving the responses of 11 students and one instructor to three alternative designs lead to the final working design consisting of two separate graphs to display the same data by different means (Fig. 1): (A) an abstracted Topological View displays the media type and relative (not absolute) temporal location and relationship of topics in the video; temporal discontinuities within a topic are illustrated by tapering connecting lines; (B) a thumbnailed Key Frame View facilitates access to full-size key frames (and the video itself) within each topic. Key frames are further distinguished by their media types. We have identified six: board, class, computer, illustration, podium, and sheet.

## 2. CLASSIFICATION BY MEDIA TYPE

The first step in the segmentation process is to assign each key frame to a media type. This classification uses a decision tree of static image feature filters (Fig. 3), such as color information in certain spatial arrangements, color patterns, and features such as edge information.

Key frames (other than already labeled ppt) are first analyzed for characteristic visual features: they are padded with a 5 to 10% border of black pixels, and are among the darkest key frames. Classifying black bordered or dark frames as computer frames was found to be 100% accurate.

Remaining key frames (which may include additional computer frames) are analyzed by color content. Key frames with mostly green color are labeled as candidates for media types board and podium; those with mostly white color for computer and sheet; and any remaining are candidates for a more complex analysis for 4 media types.

The semantic difference between the board and podium media types is derived from the behavior of the instructor. When the instructor is using the blackboard, the cameraman tends to focus on the blackboard. If the instructor is interacting with the class, the camera tends to focus on the instructor. These podium key frames contain portions of green color concentrated in vertically central regions of the image, with bottom portions of the image colored differently (Fig. 2b). Predominantly green key frames lacking green color in their bottom 10% are classified as podium. In contrast, board key frames empirically are observed to be of two major kinds. The first is a large green area that includes a green lower border. The second is a smaller green area (due to occlusion by the instructor) that still has a nearly complete green border on all sides. Board and podium key frames are accurately classified at this stage 97% of the time. Errors consist of predominantly green computer or illustration key frames.

Candidates for the computer media type, which at this stage are primarily white, are distinguished by the presence of the horizontal linearity of their contents: rectangular imagery, tables, menu bars, etc. By using a Laplacian edge detector, we extract an edge image, and compute from it a weighted measure of the presence of horizontal lines, giving exponentially more weight to longer lines. Key frames with a measure above a threshold are classified as computer frames, with 99.9% accuracy.

Remaining candidates are classified as sheet media type if they include a sufficiently large amount of white, light gray, and/or skin tones, with 100% accuracy.

Key frames without dominant green or white regions can still be classified as any media type other than sheet. This last stage handles, for example, zoomed-out frames of the board or podium, or computer frames and illustrations having colors other than white or green. An empirically derived sequence of tests revisits the computer, podium, and board media types, leaving the illustration media type to be the default classification if the other three types fail.

The key frame is first tested for media type computer by computing its horizontal line measure, and a related measure of vertical or horizontal color repetition. Frames exceeding thresholds in either measure are classified as computer. Frames not meeting these conditions are reexamined for the heuristic features specified for board and podium given above, with similar classification accuracy. Failure of these tests results in the key frame being classified as media type illustration.

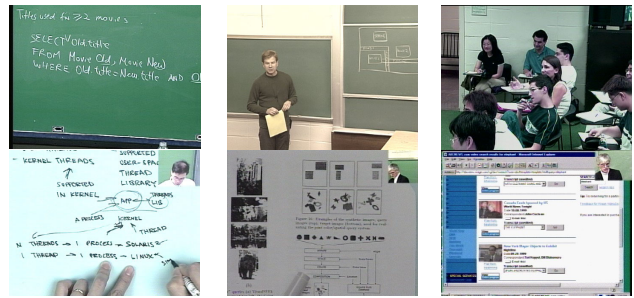

**Figure 2.** Examples from each of the six media types collected from videos of 5 courses: board, podium, class, sheet, illustration, computer.

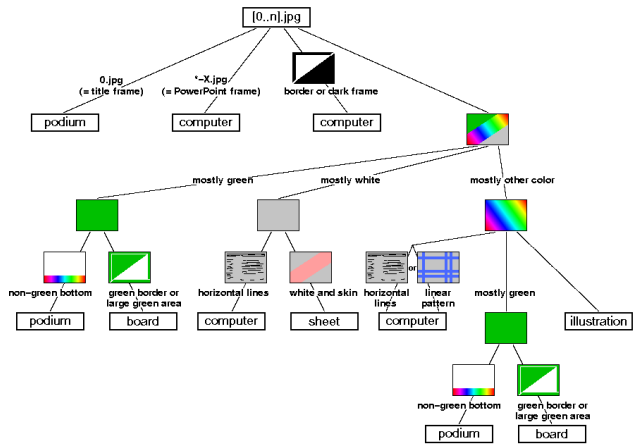

**Figure 3.** Visual feature filters for classification by media type.

Illustration key frames are mislabeled as computer frames about 23% of the time. Illustrations are typically extracted from printed media that exhibit horizontal linearity and some color repetition.

## 3. TOPOLOGICAL SEGMENTATION

The proprietary software that selects key frames does so in a way that is mostly sensitive to instructor motion, concentrating its captures during periods of relative visual calm. Consequently, the key frames are highly redundant. Therefore, we cluster similar key frames into topics based on visual content whenever the more recent frame elaborates on the visual information found in the more distant one. This clustering reduces a set of key frames to a set of clusters (topics) of similar key frames that is 5 to 20% of that size (Fig. 4).

We apply topological clustering only to board and sheet key frames, because these are the most informative media types. First, images are filtered to extract writing. Secondly, images of this writing are matched using selected sub-windows of content.

### 3.1. Filtering

The essential distinctions between media types has an operational consequence: different sets of filters are applied to either type. For board key frames, this means filtering

$X^1$ $X^2$ $Y^6$ **A**$^1$ $X^{10}$ **B**$^8$ $X^3$ $Y^1$ $Y^1$ **C**$^{29}$ $X^{12}$ **D**$^{11}$ **E**$^{21}$ **F**$^{15}$ **G**$^{28}$ $X^{16}$ $Y^1$ **H**$^7$ **I**$^{42}$ $X^4$ **I**$^5$
$X^1$ **H**$^2$ **I**$^8$ $X^{10}$ **J**$^{14}$ $X^1$ **K**$^7$ **J**$^1$ **K**$^6$ **J**$^2$ **K**$^{11}$ **J**$^5$ $X^6$ **J**$^5$ **H**$^3$ **I**$^1$ $X^1$ $Y^{13}$ $X^1$ $X^1$

**Figure 4.** For the video displayed in Figure 1, 323 key frames have been reduced to 41 clusters. *X* and *Y* denote podium and computer key frames, and letters **A** though **K** distinct clusters of board key frames. The exponents denote the number of similar frames in a contiguous sequence.

out the board and all surrounding artifacts, while sharpening the chalk marks. For sheet key frames, a filter removes all areas that do not relate to the material written on the lighter background.

The board filter addresses the problem of poor board versus chalk contrast. First, potential blackboard pixels are isolated with a simple green color filter 5A(b). Using the edges found by the edge filter in 5A(f) and the potential blackboard pixels from 5A(b), the board in 5A(a) is flooded to obtain the largest closed blackboard region(s) in 5A(c). Because the result excludes any writing, the result is outlined in 5A(d) and flooded again in 5A(e).

The foreground is extracted beginning with a 3x3 Laplacian edge filter in 5A(f). Edge artifacts from the borders of homogeneous regions are detected by a horizontal and vertical color similarity filter and removed in 5A(g). A morphological filter in 5A(h) removes noise while restoring content pixels mistakenly removed from 5A(f). A second morphological filter in 5A(i) restores content pixels mistakenly removed from 5A(a). The foreground pixels are ANDed with background pixels to recover writing pixels on the board. Finally, large blobs of writing are removed in 5A(k) because they are not useful in matching. We call this binary frame the derived content frame.

The sheet filter uses the same methodology for extracting writing from sheets of paper. However, because ink on fresh sheets of paper usually has higher contrast than chalk on erased blackboards, no noise filtering is necessary for the foreground, and hence stages 5A(g – i) are omitted.

### 3.2. Matching

Matching the content between two key frames is complicated by the three degrees of freedom allowed to the otherwise fixed cameras: tilt, pan, and zoom. These introduce translation, scale, and perspective changes. Perspective is handled by an implicit para-perspective method: both frames are considered to be made up of small local windows. As few scale-invariant features are expected, scale is explicitly modeled by successively rescaling one of the pair by a range of 14 scaling factors (from 0.6 to 1.7) experimentally derived.

Given two key frames *i* and *j*, with *j* the more recent, we extract a set of features in *i* by means of an interest operator, and find their correspondences in *j*. If sufficient similarity exists, frame *j* is considered to be an elaboration of the topic in frame *i*.

Features in frame *i* are extracted in the form of up to 6

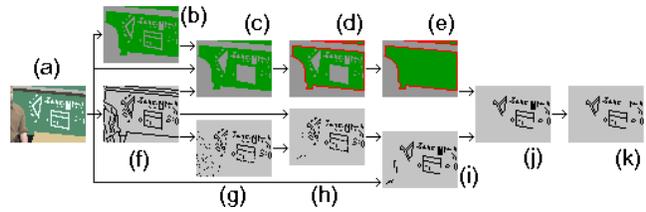

**Figure 5.** Flow diagram of the board filtering process. (a) original image, (b) green color filter, (c) flooded board, (d) outline of flooded area, (e) complete flooding of outlined area (board), (f) edge filter, (g) horizontal and vertical color similarity filter, (h,i) morphological filters, (j) ANDed combination of (e) and (i) results in extraction of board contents, (k) large pixel blob filter removes useless features.

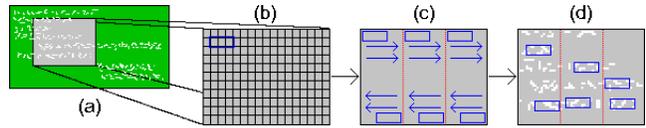

**Figure 6.** (a) A board key frame may only contain a portion of the full blackboard. (b) Content pixels from the filtered image are extracted from fixed position in a superimposed grid. (c,d) Up to 6 interesting sub-windows are identified in 3 equally sized vertical strips over the grid.

identically sized, wide aspect ratio sub-windows of interest (Fig. 6). Their selection proceeds as follows:

1. Divide the derived content frame into three equal vertical strips. 6A(c)
2. For each strip, scan the interest window over a coarse grid of locations in the image. 6A(b,c)
3. Count the content pixels *cc* in window placement. If $low \leq cc \leq high$, stop and report window position. 6A(d)
4. Repeat 2 and 3, except scanning bottom-up.
5. If both reports are the same, keep only one.

This heuristic search reflects the empirical observations of both the units of writing and the camera motions observed in the videos. Since panning for boards dominates tilting for sheets, the search enforces a horizontally balanced window acquisition. Windows are sized so that their height roughly corresponds to two lines of text, while their width is about twice that size to reflect the lengths of average words.

Empirically, it is observed that windows of interest contain between 5 to 30% content pixels. This range provides enough pixels to match with, but not so many as to prohibit the creation of distinctive configurations of pixels.

Having found windows in frame *i* (Fig. 7(a)), to increase match likelihood we next blur the derived content frame *j* by opening it with a 3x3 mask. We now find the best correspondences to these windows in the blurred derived content frame *j*, and compute a total match score:

1. Find the best location of each sub-window from *i* in *j* in the usual manner of template matching. 7(b)
2. Define image match quality for each sub-window match as the amount of matched writing divided by the total amount of writing. 7(c)
3. Define consistency of translation of the windows as the negative standard deviation of the lengths of the translation vectors for each sub-window pair. 7(d)

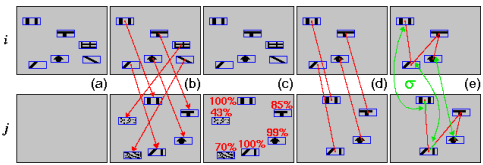

**Figure 7.** Matching algorithm. (a) find interesting sub-windows in i; (b) find corresponding sub-windows in j; (c) determine image match quality; (d) compute σ of translation vectors; (e) compute σ of change of intra-window distances.

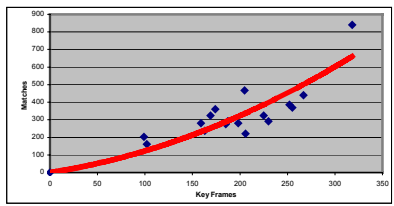

**Figure 8.** Quadratic regression of number of matches versus number of key frames over 17 different videos: M(f) = .84f + .0039f².

| N = 4479 | Classified | | | | |
|---|---|---|---|---|---|
| | B | P | S | I | C |
| B (696) | - | 0 | 0 | 0 | 0 |
| P (431) | 0 | - | 2 | 9 | 0 |
| S (2708) | 0 | 2 | - | 0 | 2 |
| I (387) | 0 | 0 | 0 | - | 89 |
| C (257) | 0 | 9 | 0 | 0 | - |

**Table 1.** Confusion Matrix for (B)oard, (P)odium, (S)heet, (I)llustration, (C)omputer: number of key frames that were incorrectly classified.

4. Define consistency of spatial arrangement as the negative of the standard deviation of the errors between corresponding intra-window distances. 7(e)
5. Total match score is a weighted sum of the number of windows in the match, image match quality, translation consistency, and spatial arrangement consistency.

We define a topic to be a temporally ordered but possibly non-consecutive sequence of key frames. Topics are themselves temporally ordered by the time of their most recent frame. The key frames of the video can now be clustered into topics by having each successive key frame either extend an existing topic sequence or start a new one:

1. The first key frame of the video forms the first topic.
2. Each succeeding key frame of the video is matched to the most recent frame of the most recent topic.
3. If this match succeeds, the most recent topic is extended by the incoming frame, and the frame becomes the most recent frame of the topic.
4. If the match fails, the incoming key frame is matched in sequence to the most recent frame of the other topics, in the order of topic recency.
5. If the incoming key frame finds a match, it extends that topic, and it becomes the most recent frame of the topic, and the topic becomes the most recent topic. If no match is found at all, the incoming key frame starts a new topic, and becomes the most recent topic.

Because for media type sheet there are no erasures, the matching is performed by finding sub-windows in the older frame and searching for them in the newer one. For media type board, where erasures are common, matching can also proceed in the reverse direction as well.

## 4. COST OF MATCHING AND ACCURACY

Under reasonable assumptions, the matching algorithm is approximately linear. We assume that the number of topics grows linearly but slowly; in the 17 videos the ratio of topics to total number of key frames is small, with average 14.8/200=.074. Statistically, a match between two consecutive key frames occurs 89% of the time, a match between two non-consecutive key frames 3.6%, and no match occurs (i.e. a new topic is formed) 7.4% of the time.

The expected cost of a match at frame $f$ is composed of three terms: the expected cost of performing a match with the previous key frame (always), the expected cost of finding a match with a key frame from a prior topic (a prior topic is extended), the expected cost of performing a match with a key frame from all prior topics (new topic is started)

$Match(f) = p_{exact}*1 + p_{previous}*O(f)/2 + p_{new\ topic}*O(f)$

Empirically, there are an average of 200 key frames per board or sheet segmentation; the probability of matching with the current topic is 178/200=.89, with a topic prior to the most recent topic is 7.2/200=.036, and with no previous topic is 14.8/200=0.074. The cost for matching 200 frames becomes: $M(f) = .89f + .0034f^2$

This result is very close to the quadratic regression in Figure 8. For most videos, the quadratic term is negligible.

We have collected data over 17 extended videos measuring 40 total hours that suggests several properties of the underlying processes. Media type classification is robust, as most media types were correctly detected between 97 and 100% of the time (Table 1). The only exception is detection of type illustration. Topological Segmentation performed equally well at a success rate of more than 96%.

Future investigations would include more careful optimizations of the match, particularly with regard to scale, and the automatic extraction of significant terms or diagrams from repeated key frames to augment the Topological View with symbolic index terms.